
\input harvmac.tex
%
%
%
%
%
\baselineskip=12pt
\lref \rAkinetic { D. Arnaudon, {\sl Composition of kinetic momenta:
the ${\cal U}_q(sl(2))$ case,}
preprint CERN-TH.6730/92, to appear in Commun. Math. Phys.}
\lref \rDK {  C. De Concini and V.G. Kac,
{\sl Representations of quantum groups at roots of 1,}
Progress in Math. {\bf 92} (1990) 471 (Birkh\"auser).}
\lref \rDKP {  C. De Concini, V.G. Kac and C. Procesi,
{\sl Quantum coadjoint action,}
Preprint Pisa (1991).}
\lref \rFRT { L. D. Faddeev, N.Yu. Reshetikhin and L.A.
Takhtajan, {\sl
Quantization of Lie groups and Lie algebras}, Leningrad Math.
J.  {\bf 1} (1990) 193.}
\lref \rKerDipl { T. Kerler,
{\sl Darstellungen der Quantengruppen und Anwendungen,}
Diplomarbeit, ETH--Zurich,
 August 1989. }
\lref \rRosB { M. Rosso, {\sl An analogue of the P.B.W.
theorem and the universal ${\cal R}$-matrix for  $\cU _h sl(N+1)$,}
Commun. Math. Phys. {\bf 124} (1989) 307.}
\lref \rRosHC { M. Rosso,
{\sl Analogues de la forme de Killing et du th\'eor\`eme
d'Harish--Chandra pour les groupes quantiques,}
Ann. Scient. Ec. Norm. Sup., $4^{\rm e}$ s\'erie, t. 23 (1990) 445.}
\def\cC{{\cal C}}
\def\cN{{\cal N}}
\def\cS{{\cal S}}
\def\cU{{\cal U}}
\def\uqn{{\cal U}_q\left(sl(N)\right)}
\def\frac#1#2{{{#1}\over{#2}}}
%
\line{\hfil                                                   CERN-TH.7025/93}
\line{\hfil                                                   ENSLAPP-A-437/93}
\Title{}{\vbox{\centerline{
Polynomial Relations in the Centre of ${\cal U}_q(sl(N))$
}}}

\centerline{Daniel Arnaudon $^{a,b,}$
\footnote{$^{1}$}
{arnaudon@surya11.cern.ch, arnaudon@lapphp1.in2p3.fr}
\footnote{}{Address after $1^{\rm st}$ Oct. 1993: ENSLAPP
}
and Michel Bauer $^{a,}$
\footnote{$^{2}$}{mbauer@surya11.cern.ch}
\footnote{}{On leave from SPhT, CE Saclay, 91191 Gif-sur-Yvette
Cedex, France}
}
\vfill
\centerline{$^a$ Theory Division, CERN, 1211 Geneva 23, Switzerland}
\medskip
\centerline{$^b$ ENSLAPP\footnote{$^3$}{
URA 14-36 du CNRS, associ\'ee \`a l'E.N.S. de Lyon et au L.A.P.P.
d'Annecy-le-Vieux.},
Chemin de Bellevue BP 110, 74941
Annecy-le-Vieux Cedex, France}

\vfill
When the parameter of deformation $q$ is a root of unity,
the centre of  ${\cal U}_q(sl(N))$ contains, besides the usual
$q$-deformed Casimirs, a set of new generators, which are basically the
$m$-th powers of all the Cartan generators of ${\cal U}_q(sl(N))$.
All these central elements are however not independent.
In this letter, generalising the well-known case of ${\cal
U}_q(sl(2))$, we
explicitly write polynomial relations satisfied by the generators
of the centre. Application to the parametrization of irreducible
representations and to fusion rules are sketched.

\vfill\vfill

\leftline{CERN-TH.7025/93}
\leftline{October 1993}
\leftline{hep-th/9310030}

\Date{}

\newsec{Symmetric polynomials}
In the course of our study of Casimirs for ${\cal U}_q(sl(N))$ we shall
repeatedly encounter special symmetric polynomials of $N$ variables denoted
by $x_1,\cdots,x_N$.
In the classical
theory of Casimirs of $sl(N)$,
symmetric polynomials in $N$ variables play a special role since
the Weyl group of $sl(N)$ is the symmetric group on
$N$ objects. The Weyl group acts on the Cartan torus and on its Lie
algebra $\Im$
and a well-known theorem of Harish--Chandra says that there is natural
isomorphism between the centre of ${\cal U}(sl(N))$ and the Weyl-invariant
elements of ${\cal U}(\Im)$. The more precise corresponding statements in the
case of ${\cal U}_q(sl(N))$ when $q$ is a root of unity will be given below.

The elementary symmetric polynomials $c_1, \cdots,c_N$ are defined by
the identity
\eqn\eG{\prod_{i=1}^N (1-tx_i)=1-c_1t+c_2t^2-\cdots+(-1)^Nc_Nt^N
\equiv G(t).}
Hence for $i=1,\cdots,N$
\eqn\ec{c_i=\sum_{1 \leq j_1 < \cdots < j_i \leq N} x_{j_1} \cdots
x_{j_i}}
and it is an old theorem attributed to Newton that any symmetric polynomial
in $x_1,\cdots,x_N$ (with coefficients in a ring) is a polynomial in $c_1,
\cdots,c_N$ with coefficients in the same ring.

The polynomials of interest to us in the sequel are generalisations of the
elementary ones obtained by replacing the variables $x_i$ by their
$m^{\rm {th}}$
power. Hence we define $P^{(N)}_{i,m}(c_1,\cdots,c_N)$ for $i=1,\cdots,N$ and
$m=1,2,\cdots$ by the
identity
\eqn\eP{\prod_{i=1}^N
(1-tx_i^m) \equiv 1-P^{(N)}_{1,m}t+P^{(N)}_{2,m}t^2
-\cdots+(-1)^NP^{(N)}_{N,m}t^N.}

It is useful to have expressions displaying
these  polynomials directly
in terms of the elementary symmetric polynomials $c_i$ (and not in terms of
the variables $x_1,\cdots,x_N$).
A method that works nicely
for fixed $m$ is to remark that for any primitive $m^{\rm {th}}$ root
of \hbox{unity $q$}
\eqn\eroot{1-t^mx_i^m=\prod_{l=1}^m (1-q^l tx_i)\;,}
from which we deduce that
\eqn\eprodG{\prod_{i=1}^N (1-t^mx_i^m)=\prod_{l=1}^m G(q^l t)\;.}
Finally we obtain the desired result
\eqn\eprodGii{1-P^{(N)}_{1,m}t^m+P^{(N)}_{2,m}t^{2m}
-\cdots+(-1)^NP^{(N)}_{N,m}t^{Nm}=\prod_{l=1}^m
G(q^l t).}
This formula makes the computation of the $P^{(N)}_{i,m}$'s for
reasonnable values
of $N$ and $m$ tractable, at least with the help of a computer.

The polynomials $P^{(N)}_{1,m}$ will play a distinguished role in what
follows. The
generating function
\eqn\egen{\sum_{m=1}^{\infty} P^{(N)}_{1,m}{{t^m}\over{m}}}
is easy to express in terms of $c_1, \cdots, c_N$, because
\eqn\egenii{-\log (1-tx_i)=\sum_{m=1}^{\infty}x_i^m {{t^m}\over{m}}\;,}
leading to
\eqn\egenG{\sum_{m=1}^{\infty} P^{(N)}_{1,m}{{t^m}\over{m}}=-\log G(t)\;.}

Let us end this section with some examples of these polymials. First note
that
for our purpose we will have to consider only the particular case $c_N=1$.

In the case of ${\cal U}_q(sl(2))$, we will need
$P^{(2)}_{1,m}$, which is closely related to
the $m^{\rm {th}}$
Chebichev polynomial of the first kind.

In the case of ${\cal U}_q(sl(3))$ and $m=5$, the polynomials of
interest are
\eqn\ePthree{\eqalign{
P^{(3)}_{1,5}(c_1,c_2) = & \ c_1^5 - 5 c_1^3 c_2 + 5 c_1 c_2^2 + 5 c_1^2
- 5 c_2 \cr
P^{(3)}_{2,5}(c_1,c_2) = & \ c_2^5 - 5 c_1 c_2^3 + 5 c_1^2 c_2 + 5 c_2^2
- 5 c_1 \;.\cr}}

In the case of ${\cal U}_q(sl(4))$ and $m=5$, we will need
\eqn\ePfour{\eqalign{
P^{(4)}_{1,5}(c_1,c_2,c_3) = & \
c_1^5 - 5 c_1^3 c_2 + 5 c_1 c_2^2 + 5 c_1^2 c_3 - 5 c_2 c_3 - 5 c_1 \cr
P^{(4)}_{2,5}(c_1,c_2,c_3) = &  \
c_2^5 - 5 c_1 c_2^3 c_3 + 5 c_1^2 c_2 c_3^2 + 5 c_2^2 c_3^2
- 5 c_1 c_3^3 + 5 c_1^2 c_2^2 \cr
& - 5 c_2^3 - 5 c_1^3 c_3 - 5 c_1 c_2 c_3
+ 5 c_3^2 + 5 c_1^2 + 5 c_2
\cr
P^{(4)}_{3,5}(c_1,c_2,c_3) = & \
c_3^5 - 5 c_2 c_3^3 + 5 c_2^2 c_3 + 5 c_1 c_3^2 - 5 c_1 c_2 - 5 c_3 \;.\cr}}

\newsec{$\uqn$ at roots of unity}
Let $\{ \alpha_1,...,\alpha_{N-1} \}$ be the set of simple roots of
$sl(N)$.
We define vectors $\epsilon_1$,..., $\epsilon_N$ by
$\alpha_i=\epsilon_i-\epsilon_{i+1}$ and
$\sum_{i=1}^N \epsilon_i = 0$.

The ``simply connected''
quantum group $\uqn$ is defined by the generators
$e_{i}$, and $f_{i}$, for $i=1,...,N-1$,
and $k_{\pm \epsilon_i}$ for $i=1,...,N$,
and the relations
\eqn\erelations{
\left\{ \eqalign{
  k_{\beta_1} k_{\beta_2} &= k_{\beta_1 +\beta_2} \; ,
\cr
  k_{\epsilon_i}  e_{j}  k_{\epsilon_i}^{-1}
&= q^{\delta_{ij} -\delta_{i-1, j}} e_{j} \;,
\cr
  k_{\epsilon_i}  f_{j}  k_{\epsilon_i}^{-1}
&= q^{-\delta_{ij} +\delta_{i-1, j}} f_{j}  \;  ,
\cr
  [e_{i},f_{j}] &= \delta _{ij} {k_{\alpha_i} - k_{\alpha_i}^{-1}
\over q -q^{-1}}\; ,
\cr
  [{e}_{i},{e}_{j}] &=[{f}_{i},{f}_{j}] =0 \qquad {\rm for} \qquad
            \left|{i-j}\right|\ge 2 \;,
\cr
  e_{i}^{2} e_{i\pm 1}- & (q+q^{-1})  e_{i} e_{i\pm 1} e_{i}
     +e_{i\pm 1} e_{i}^{2}   =0\; ,
\cr
  f_{i}^{2} f_{i\pm 1}- & (q+q^{-1})  f_{i} f_{i\pm 1} f_{i}
     +f_{i\pm 1} f_{i}^{2}   =0\;.
\cr   } \right. }

Let $\cU^0$ be the subalgebra generated by the $k_{\epsilon_i}$'s, and
$\cU^+$, $\cU^-$ the subalgebras generated by the $e_i$'s, $f_i$'s,
respectively.

Two sets of quantum analogues of the roots vectors
are inductively defined as
\eqn\eeij{
\cases{
e_{i,i+1} = \tilde e_{i,i+1} \equiv e_i & for $i=1,...,N-1$ \cr
e_{i,j+1} = e_{ij}e_{j}-q^{-1}e_{j}e_{ij} & for $i<j$  \cr
\tilde e_{i,j+1} = \tilde e_{ij}e_{j}-q e_{j}\tilde e_{ij} & for $i<j$ ;  \cr
}}
as are the $f_{ij}$ and $\tilde f_{ij}$.

Quantum analogues of
Poincar\'e--Birkhoff--Witt bases can
be built with ordered monomials in these generators \rRosB.

\medskip
When $q$ is not a root of unity, there exists a quantum analogue of
Harish--Chandra theorem \refs{\rRosHC, \rDKP}: there exists an algebra
isomorphism
$h$ from $Z$, the centre of $\uqn$, to the algebra of
symmetric polynomials in the $k_{2\epsilon_i}$.
This isomorphism $h$ can be written as $h=\gamma^{-1} \circ h'$,
with the following notations:
$h'$ is the projection on $\cU^0$, within the direct sum
$\cU=\cU^0 \oplus (\cU^-\cU + \cU \cU^+)$, with $\cU\equiv\uqn$;
$\gamma$ is the
automorphism of $\cU^0$ given by
$\gamma(k_{2\epsilon_i}) = q^{N+1-2i} k_{2\epsilon_i} $.

A set of generators of $Z$ is given by
\eqn\eCici{\{\cC _i=h^{-1}(c_i(k_{2\epsilon_1},...,
   k_{2\epsilon_N}))\}_ {i=1,...,N-1}.}
An expanded expression for these generators (denoted
there by $\tilde c_k$) appears in \rFRT\
in the form (up to slight changes of
convention and normalization):
\eqn\eCi{\cC _i = q^{i(N-i)}\cN_i(q^{-2})^{-1}\cN_{N-i}(q^{-2})^{-1}
\sum_{\sigma,\sigma'\in \cS(N)}
(-q^{-1})^{l(\sigma)+l(\sigma')}
l^{(+)}_{\sigma_{1}\sigma'_{1}}...
l^{(+)}_{\sigma_{i}\sigma'_{i}}
l^{(-)}_{\sigma_{i+1}\sigma'_{i+1}}...
l^{(-)}_{\sigma_{N}\sigma'_{N}}\;,
}
where $\cN_i(x)=\prod_{n=1}^i (1+\cdots+x^{n-1})$,
where $l(\sigma)$ is the length of the shortest expression of the
permutation $\sigma$ in terms of simple transpositions, and where
$$\eqalign{l^{(+)}_{ii} &= \left( l^{(-)}_{ii} \right)^{-1}
 = k_{\epsilon_i} \cr
l^{(+)}_{ij} &= l^{(-)}_{ji} =0 \qquad \hbox{for} \quad i>j \cr
l^{(+)}_{ij} &= (q-q^{-1}) (-1)^{j-i+1} \tilde f_{ij}  k_{\epsilon_i}
\qquad \hbox{for} \quad i<j \cr
l^{(-)}_{ij} &= (q-q^{-1}) (-1)^{j-i} k_{-\epsilon_i} \tilde e_{ij}
\qquad \hbox{for} \quad i>j \;.\cr}$$

The first and last of these Casimirs are explicitly given by
\eqn\eCun{
\cC _1 =    \sum_{i=1}^N q^{N+1-2i} k_{2\epsilon_i}
         + (q-q^{-1})^2 \sum_{1\le i<j \le N}(-1)^{j-i-1}
          q^{N+1-i-j} \tilde f_{ij} e_{ij}  k_{\epsilon_i +\epsilon_j}
}
and
\eqn\eCN{
\cC _{N-1} = \sum_{i=1}^N q^{-N-1+2i} k_{-2\epsilon_i}
         + (q-q^{-1})^2 \sum_{1\le i<j \le N}(-1)^{j-i-1}
         q^{-N-1+i+j}  f_{ij} \tilde e_{ij} k_{-\epsilon_i -\epsilon_j}\;.
}

When $q$ is a root of unity, the image $Z_1$ of $h$ is still a
well-defined central subalgebra of $\uqn$ \rDKP, but it does not
generate the whole centre. Let $Z_0$ be the subalgebra of $\uqn$
generated by the elements $f_{ij}^m$, $e_{ij}^m$ and
$k_{m\epsilon_i}$. (We could also replace
$f_{ij}$ by $\tilde f_{ij}$, or $e_{ij}$ by $\tilde e_{ij}$, this
would lead to the same $Z_0$.) When $m'$ is odd, these elements are
central, and the centre $Z$ of $\uqn$ is actually generated by $Z_0$
and $Z_1$ \rDKP.

\newsec{Relations in the centre of $\uqn$}
{\bf Theorem: }{\sl If $m'$ is odd,
the following relations are satisfied in the centre of $\uqn$,
\eqn\eRi{
\eqalign{
P_{1,m}^{(N)} &  (\cC _1,...,\cC _{N-1}) =
         \sum_{i=1}^N q^{m(N+1)} k_{2m\epsilon_i} \cr
     &    + (q-q^{-1})^{2m}   \sum_{1\le i<j \le N}(-1)^{m(j-i-1)}
       q^{m(N+1-i-j)} \tilde f_{ij}^m e_{ij}^m  k_{m\epsilon_i +m\epsilon_j}
   \cr}
}
and
\eqn\eRN{
\eqalign{
P_{N-1,m}^{(N)} & (\cC _1,...,\cC _{N-1}) =
          \sum_{i=1}^N q^{-m(N+1)} k_{-2m\epsilon_i} \cr
      &   + (q-q^{-1})^{2m}   \sum_{1\le i<j \le N}(-1)^{m(j-i-1)}
      q^{m(-N-1+i+j)}  f_{ij}^m \tilde e_{ij}^m k_{-m\epsilon_i -m\epsilon_j}
   \cr}
}
}

{\it Remark 1:}
Actually, all the powers of $q$ are equal to $1$ since $m'$ is odd,
but we conjecture that these formulae remain true for even $m'$.
[In this case, the terms $\tilde f_{ij}^m$, $e_{ij}^m$ and
$k_{m\epsilon_i +m\epsilon_j} $ are not individually central, but
their products are.]

{\it Remark 2:}
To get the right-hand sides of these relations, one simply replaces
each term (including numerical factors) in the expression of $\cC_1$
(resp. $\cC_{N-1}$) by its $m^{\rm th}$ power.
This remarkable  relationship seems to hold between
$P_{i,m}^{(N)} (\cC _1,...,\cC _{N-1})$ and $\cC_i$ for the other
values of $i$ as well, if $\cC_i$ is written in a suitable
Poincar\'e--Birkhoff--Witt basis.

{\it Proof of the theorem:}
\item{\it a. }{We first apply the relations \erelations\ and
\eeij\ in order to write \eRi\ and \eRN\ and the $\cC_i$'s
in the Poincar\'e--Birkhoff--Witt  basis. Then
\eqn\eproof{\eqalign{
h\left( P_{1,m}^{(N)}   (\cC _1,...,\cC _{N-1}) \right) &=
P_{1,m}^{(N)}   (h(\cC _1),...,h(\cC _{N-1}))       \cr
 &=  \sum_{i=1}^N q^{m(N+1)} k_{2m\epsilon_i} \cr  }}
(and the corresponding formula with $P_{N-1,m}^{(N)}$).
This follows  from the
definitions of the first section. It then appears
that this projection belongs to $Z_0$,
and hence so does the whole result (\rDKP\ Prop. 6.3.c).
This part of the proof also applies to
$P_{i,m}^{(N)}   (\cC _1,...,\cC _{N-1})$ for $1<i<N-1$, whereas the
second part is limited to the cases $i=1$ or $i=N-1$.
}
\item{\it b. }{We can then use considerations on the degrees of the
monomials appearing in $P_{1,m}^{(N)}$ (and $P_{N-1,m}^{(N)}$) to
complete the proof. The
term of highest degree
of $P_{1,m}^{(N)}$ (resp. $P_{N-1,m}^{(N)}$)
is indeed $\cC_1^m$ (resp. $\cC_{N-1}^m$), and it is
also the only term of degree $m$. According to the form of the $\cC_i$
\eCi, only monomials of degree at least equal to $m$ can contribute to
non-trivial terms belonging to $Z_0$: a necessary condition is indeed
that the products of root vectors they contain correspond to an
element of the root lattice $R$ belonging to $mR$. For the same
reason, the contribution of the monomial of degree $m$  is precisely
the second part of the right-hand side of \eRi\ (resp. \eRN).
}
\medskip
Relations \eRi, \eRN\ differ, for $N>2$,
from the equation in the last remark of \rDKP. In particular, the
degree of the polynomial is different.
In the case of $\cU_q(sl(2))$, the relation \eRi\ was already given in
\rKerDipl.

\newsec{Applications}
\item{\it a.} Parametrization of generic  irreducible representations:
\hfil\break
We know from \rDK\ that generic irreducible representations of $\uqn$
are characterized by the values of the central elements on them. Once
the values of the elements of $Z_0$ are determined, a choice between
$m^{N-1}$ values for $\cC_1,...,\cC_N$ remain. A nice way to
parametrize them is to write, for a representation $\rho$,
\eqn\eCizeta{\rho\left(\cC_i\right) =
c_i(\zeta_1,...,\zeta_N)}
with $c_i$ defined in \ec\ and
$\prod_{i=1}^N \zeta_i = 1$. (Note the absence of $h^{-1}$, by
comparison with \eCici.) The $m^{N-1}$ irreducible representations on
which the elements of $Z_0$ take the same value simply correspond to
the parameters
\eqn\eallzeta{q^{p_1}\zeta_1,...,q^{p_N}\zeta_N,}
with
$p_1,...,p_N\in {\bf Z}$  and $\sum_1^N p_i = 0 \hbox{ mod } m$.
Since
\eqn\ePCcz{\rho\left( P_{i,m}^{(N)} (\cC _1,...,\cC _{N-1}) \right)
   = c_i(\zeta_1^m,...,\zeta_N^m) }
for $1\le i \le N-1$, these
sets of parameters indeed correspond to the sets of solutions for the
$\cC_i$'s,  to the system of $N-1$ equations
including \eRi\ and \eRN.
\hfil\break
With this parametrization, the $\zeta_i$ become powers of
$q$ when the central elements $e_{ij}^m$, $f_{ij}^m$ and
$k_{2m\epsilon_i}$  take the values $0$, $0$ and $1$ respectively.
In
this highly non-generic case, a finite number of irreducible
representations is related to the same parametrization.
These representations are  $q$-deformations of
classical representations.
\medskip
\item{\it b.} Application to fusion rules:
\hfil\break
We suggest that these relations and the above parametrization could
help in the study of fusion of
unrestricted (generic) irreducible representations of $\uqn$, as
in \rAkinetic\ in the case of $\cU(sl(2))$. The strategy would be the
following: to evaluate the values of the elements of $Z_0$ in the
tensor product of two irreducible representations (they are scalar);
find then a solution for the parameters $\zeta_i$ compatible with
these values. Then all the irreducible representations characterized
by the parameters \eallzeta\ should appear in the fusion rule, with
multiplicity $1$ in the case generic$\otimes$minimal--periodic, and with
multiplicity $m^{(N-1)(N-2)/2}$ in the case generic$\otimes$generic.

\bigskip
{\bf Ackowledgements:} Some preliminary computations were done using the
program FORM, by Jos Vermaseren. We thank Jean Orloff for his advices
on symbolic manipulation programs. We also thank Alain Lascoux for
an illuminating discussion on combinatorics and for various comments.
\listrefs
\bye